\documentstyle[11pt,paspconf]{article}

\input{psfig}

\begin{document}

\title{Black Hole Event Horizons and X-ray Nova Luminosities - Update}

\author{M.R. Garcia, J.E. McClintock, R. Narayan}
\affil{Smithsonian Astrophysical Observatory}
\author{P. J. Callanan}
\affil{Univ. College, Cork}

\section{Introduction}
Recently, Narayan, Garcia and McClintock (NGM97$^1$) showed that there
is a difference in the {\bf quiescent} luminosities of black hole and
neutron star x-ray nova: when the quiescent, Eddington scaled
luminosities are compared, every black hole x-ray nova (BHXN) is
fainter than every the neutron star x-ray nova (NSXN).  To this
earlier work we add three more BHXN, \hbox{GRO~J1655-40},
GRO~J0422+32, and 4U1543-47.  This increases the significance of the
difference between the quiescent luminosities to the $>$99\% level.

\section{Observations}

\indent For {\bf GRO~J0422+32}, the $3\sigma$ upper limit to the quiescent
luminosity (0.5-10~keV) is from a 19~ks ROSAT HRI observation
extracted from the HEASARC public archive.  In order to compute the
luminosity we have assumed a power-law spectrum with $\alpha_N = 2.1$,
as has been observed from V404~Cyg$^2$, and $log(N_H)=21.3$.  The
outburst luminosity is from Barrett, McClintock and Grindlay
1996$^{3}$, and covers the 1-200~keV band.

For {\bf GRO~J1655-40}, the ASCA SIS counting rate observed during
March 1996$^{3,4}$ indicates a luminosity of $2.5\times
10^{32}$ergs~s$^{-1}$ (0.5-10~keV), assuming $\alpha=2.1$ and
$log(N_H)=21.8$.  The outburst of July 1995 was observed to have a
luminosity of $\sim 10^{38}$~erg~s$^{-1}$ (1~keV-2~MeV), but based on
the BATSE counting rates the initial outburst may have been 2-5 times
brighter$^{6}$.  We therefore assume a maximum luminosity of $3 \times
10^{38}$~erg~s$^{-1}$.

Recent dynamical measurements of \hbox{{\bf 4U1543-47}} 
imply that it contains a black hole$^7$.  The brightest of the three outbursts
recorded from this soft transient occurred in 1983$^8$, 
and was observed with Tenma$^9$ and EXOSAT$^{10}$.
While the EXOSAT observations occurred near the peak, the Tenma
lightcurve and a comparison of the measured fluxes indicates the
source may have already decayed by a factor $\sim 3$ by the time of
the EXOSAT observations. 
Two spectral fits
to the Tenma data$^9$ give $N_H$
much higher than that found by EXOSAT and than that implied by the
optical extinction$^{11}$.   One of these fits gives $N_H =
1.6 \times 10^{22}$, at which $\sim 80\%$ of the emitted flux is detected, so
this fit may yield a reasonable measure of the luminosity.
Using these spectra parameters, we extrapolate beyond the 1.5-11~keV
Tenma band to compute a 1-20 keV luminosity of $\sim 4.0 \times
10^{39}$ at 8~kpc.
The ROSAT All-Sky Survey did not detect 4U1543-47$^{12}$, yielding a $3\sigma$
upper limit in quiescence of $5.2 \times
10^{33}$ ergs/sec (0.5-10 keV), where we have assumed $\alpha = 2.1$
and $N_H = 3.1 \times 10^{21}$cm$^{-2}$. A 20~ks ASCA observation
carried out on Feb~19~1996 may yield an even lower upper limit.

\begin{figure*}[t]
\centerline{\psfig{figure=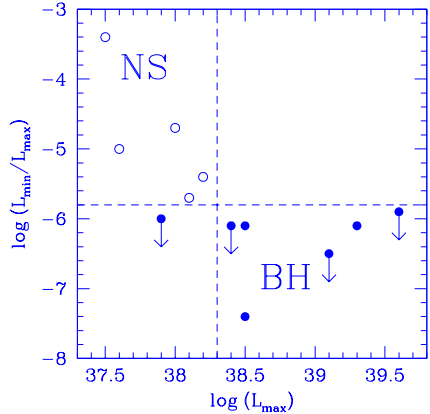,height=3.00in}}
{\small
\caption{A comparison of the maximum luminosity reached in outburst,
$L_{max}$, and the luminosity swing from outburst to quiescence,
$L_{min}/L_{max}$.  Arrows indicate upper limits.  The vertical line
is $L_{Edd}$ for a NS, the horizontal line has no physical
significance, other than to point out the dividing line between
BHXN and NSXN quiescent luminosities. }
}
\end{figure*}

\section{Conclusions}

Two separate conclusions can be drawn from the figure: {\bf 1) In
outburst}, BHXN are brighter than NSXN.  If we believe that XN
accrete at near the Eddington limit $L_{Edd}$ in outburst, then this
difference is likely due to the fact that the BH are more massive than
the NS.  GRO J0422+32 stands out as the only BHXN with $L_{max}$ below
the $L_{Edd}$ for the NSXN.  {\bf 2) In quiescence}, every BHXN is
fainter than every NSXN, when we consider the luminosity in Eddington
scaled units.  This difference is {\bf NOT} a consequence of the BHXN
higher mass and larger $L_{Edd}$: the luminosity {\bf swing} from
outburst to quiescence is higher for the BH, by a factor of $\sim 100$
on average.  This updated dataset shows a separation in luminosity
swings at the $\sim $99.5\% significance level, a factor of two higher
than that found by NGM97.  Given the reasonable assumption that 
the (Eddington scaled) mass
transfer swings in BHXN and NSXN are similar, this difference 
provides direct evidence for the existence of event horizons in the
BHXN: they have larger luminosity swings because the quiescent 
mass transfer energy is being
hidden behind the event horizon.

{\small  
\begin{table*}[t]
\begin{center}
\centerline{Luminosities of BHXN/NSXN in Quiescence and Outburst: Updated}
\vskip 10pt
\begin{tabular}{|lllrrc|}
\hline
Object & $D({\rm kpc})$ & log($N_H$) & $\log(L_{min})$ &  $\log(L_{max})$ & 
$\log(L_{min}/L_{max})$ \\
\hline
\multicolumn{6}{|c|}{Neutron Stars: NSXN}  \\
\hline
EXO0748--676 & 	10.0 	&22.35
							       &   34.1   &   	37.5   &	 --3.4 \\
Aql X--1 & 	 	2.5  	&21.32
							       &   32.6   &       37.6   &         --5.0 \\
Cen X--4 & 	 	1.2
					  	&20.85
							       &   32.4   &       38.1   &         --5.7 \\
4U2129+47 &	 	6.3  	&21.20
							       &   32.8   &	38.2   &	 --5.4 \\
H1608--522 &	3.6
					  	&22.00
							       &   33.3   &       38.0   &         --4.7 \\
\hline
\multicolumn{6}{|c|}{Black Holes: BHXN} \\
\hline
H1705--25 &  	8.6
					   	&21.44
							 	   &  $<$33.7    &     38.3     &       $<$-4.6   \\
GS2000+251 & 	2.7
					   	&21.92
							       &	$<$32.3    &     38.4     &       $<$-6.1 \\
Nov Mus 91 & 	6.5
					  	&21.21
							       &	$<$32.6    &     39.1     &       $<$-6.5 \\
A0620--00 &   	1.2
					   	&21.29
							       &	31.0     &    	38.4    &        --7.4  \\
V404 Cyg &   	3.5
					   	&22.04
							       &	33.2     &     39.3     &        --6.1 \\

4U1543-47 &	8.0$^{7}$	&22.20$^{9,11}$	& 	33.7	&	39.6	&	$<$-5.9 \\

GRO J0422+32 &	3.6$^3$		&21.3$^{13}$	&$<$31.9	&	37.9 	&   $<$-6.0 \\

GRO J1655-40 & 3.2$^{14}$		&21.8$^{15}$	&32.4	&	38.5 	&  -6.1 \\

\hline 
\end{tabular}
\end{center}
\end{table*}
}

\end{document}